\documentclass{elsart}
\usepackage{times}
\usepackage{graphicx}
\usepackage{amssymb}

\begin{document}
\begin{frontmatter}

\title{The efficiency of 
       individual optimization 
       in the conditions 
       of competitive 
       growth}

\author{J. Ko\v ci\v sov\' a $^1$}, 
\ead{\scriptsize jana.kocisova@kosice.upjs.sk}
\author{D. Horv\'ath $^{1,2,3}$}
\author{B. Brutovsk\'y $^{1}$}

\address{$^1$ Institute of Physics, 
Faculty of Science, P. J. \v{S}af\'{a}rik University, 
Park Angelinum 9,\\
041 54 Ko\v{s}ice, Slovakia}

\address{$^2$ Centre de Biophysique Mol\'eculaire, 
CNRS; Rue Charles Sadron, 45071 Orl\'eans, France}

\address{$^3$ Department of Physics, Faculty of Electrical Engineering and Informatics, 
Technical University, Letn\'a 9, 042 00 Ko\v{s}ice}

\journal{Physica A}

\begin{abstract}
The paper aims to discuss statistical properties of the 
multi-agent based model of competitive growth. Each of the agents 
is described by growth (or decay) rule of its virtual "mass" with
the rate affected by the interaction with other agents. 
The interaction depends on the strategy vector and mutual 
distance between agents and both are subjected to the agent's 
individual optimization process. Steady-state simulations 
yield phase diagrams with the high and low
competition phases (HCP and LCP, respectively) separated by critical 
point. Particular focus has been made on the indicators 
of the power-law behavior of the mass distributions with 
respect to the critical regime.  In this regime the study 
has revealed remarkable anomaly in the optimization 
efficiency. 
\end{abstract}

\end{frontmatter} 

\section{Introduction}\label{subs:intro}

Competitive growth~\cite{Cap08} is one of the most 
generic processes observable in wide range of spatiotemporal 
scales.  The rate of the growth/decay of the quantity 
$M(t)$ which represents some "virtual mass" of 
the object might be defined by the first-order 
continuous dynamics 
\begin{equation}
\frac{{\rm d} M}{{\rm d}t} = Rate(M)=  Growth(M) - Decay(M)\,.
\label{equat:logis1}
\end{equation}
Here the $Decay(M)$ which contributes into overall 
$Rate(M)$ comprises many possible processes such 
as death, destruction, slowering due to competition.  
Most of the baseline studies start from the logistic 
rate form~\cite{Mur01,Yaa08}. This minimum model 
includes the Malthusian term $Growth(M)=\alpha_{\rm m}  M$, 
$\alpha_{\rm m}>0$  and specific quadratic term 
\begin{equation}
Decay(M)=  \frac{M^2}{M_{\rm c}}\,, 
\label{equat:decay}
\end{equation}
where the competition is quantified by the 
{\em carrying capacity}~\cite{Web85}
denoted as $M_{\rm c}$.   Despite the 
logistics is well known since early 
studies of growth processes, for our purposes 
several comments regarding its structure 
will be helpful.  As $M(t)$ approaches $M_{\rm c}$ for 
$\alpha_{\rm m}>0$ the growth slows down until
$Rate(M_{\rm c})=0$. The ability to detect 
or even quantify the proximity 
of mass saturation may be invaluable in a broad  variety 
of the real world applications, enabling one to avoid 
the saturation. The rate reduction of the type 
$1- M/M_{\rm c}$ has been
used for mathematical description 
of tumor growth at tissue
level \cite{Aya06} 
as well as cellular scales \cite{Mur01,Aya06,Man03}.
At the model level, the proximity 
of carrying capacity may stimulate
growing entities to optimize 
their instant conditions. One can say that 
individual optimization~\cite{Das05,Tab94} 
raises as coevolutionary mechanism preventing 
competitors from the imposed obstructions~\cite{Cap08}. 

In the present paper we analyze statistical efficiency 
of optimization process of growing entities under the competitive 
conditions, where agent may change the strategy 
or migration routing to the more perspective regions 
by optimization. This goal requires much more detailed 
comprehensive and spatially distributed model than the 
logistic equation is, nevertheless,  simple principles 
of logistic bounding and growth can be incorporated 
into construction of multiple entities 
called autonomous agents. The agent-based paradigm~\cite{Chan06,Bon02} 
states that even relatively simple 
rules may lead to very complex emergent behavior.
Many examples  may be found, but for brevity 
we  mention only few of them: organizations of 
insect colonies~\cite{Cic01,Xia08}, 
social~\cite{Sun}, human economic behavior~\cite{Bon02} 
or firms as autonomous 
entities \cite{Coo03}.

The abstract agent-based model producing emergent "mass"
distributions is presented in this paper. 
The interpretation of  $M(t)$ or virtual 
"mass" itself depends on the scale
and application field for which the model has been suggested.
Here, the scalar can represent the variety of possible
quantities, such as length or size of an organism, 
but also diameter or volume 
of the growing bacterial colony. We assume that 
the model can be helpful also in the study of generic
features at economic 
and social scales. 
In this frame $M(t)$ may represent wealth~\cite{Bou00} 
of single seller, personal income~\cite{Aoy04}, 
money owned by citizens \cite{Chat03}, 
firm size~\cite{Fuj04}. 
From the standpoint of the total mass behavior one can, 
in principal, distinguish 
between the models that 
conserve mass~\cite{Chat03} and  
models that violate this type of invariance. 
As non-conservativeness may be understood 
as a synonym for incompleteness 
or missing information about all of the existing mass flows, 
it may be also understood as a realistic feature of 
the growth and mass-exchange models. Our focus 
on fluctuations is in part motivated by the effort 
to understand the origins of 
{\em power-law distributions}~\cite{Bou00,Aoy04,Chat03,Fuj04,Sta95,Pal07} 
in the systems which 
exhibit signs of the growth 
and competition~\cite{Cap08,Ken99}.

In the economic context the interest dates back 
to the seminal Pareto's work~\cite{Par97}. 
General formulation of the problem is motivated by the fact 
that power-law distributions also known as Zipf's law ~\cite{Dec07,Sta95} 
found also in distributions of city size \cite{Sta95}.
The power-law distribution can be identified 
in the lifetimes \cite{Rik05}, earthquake distributions \cite{Ola92}, 
firm demises~\cite{Coo03} 
or size of spatial colonies \cite{Man08} as well.
Here presented agent-based simulations contribute to the discussion
about connection of power-law distributions with critical
regime of driving parameters. 
This idea, stimulated by the concept of the feedback~\cite{Bec05} 
and self-organized criticality~\cite{Bak88,Ola92,Rik05},
has been revisited in~\cite{Hor06}. 

The structure of this paper will be as follows. 
In Section~\ref{subs:ABSmodel} we introduce 
agent-based model of the reactive interacting agents 
which manifests growth within the mass inequality constraints. 
Moreover, the agents  are able to 
optimize their strategies 
and positions~(see subsec.~\ref{subs:local_optim}). 
In Section~\ref{sec:secnum1} 
statistical characteristics obtained 
by the numerical simulations are discussed. 
These results open the question about the role of the local individual 
optimization. Finally, 
the concluding remarks 
and perspectives of our 
approach are presented.

\section{The agent-based model}~\label{subs:ABSmodel}

In the paper, we incorporate 
mechanisms of growth, individual search
and competetivness into the continuous 
stochastic agent-based 
framework and analyze equilibrium statistical consequences
of the complex model. Below we present model 
in more detailed focus now.

\subsection{State of agent}\label{subs:stateof}
The system consisting of $N$ interacting autonomous 
agents, each of them equipped with specific abilities,
is considered. It represents sufficiently
complex and general model, where the formation 
and growth phenomena are interrelated with space, 
strategic and mobility issues of the competitive world. 
At time $t$, the state of $i$th agent is described by 
the tuple $\langle {\bf X}^{(t)}_i, {\bf S}^{(t)}_i, 
M_i^{(t)}\rangle$, where ${\bf X}^{(t)}_i$ 
is the position, ${\bf S}_i^{(t)}$ the strategy and 
$M_i^{(t)}$ the mass of $i$th agent, respectively. 
The spatial coordinates are taken from real space 
\begin{equation}
{\bf X}_i^{(t)} = \left [ \, X^{(t)}_{i,1}, X^{(t)}_{i,2},
\ldots, X_{i,d_{\rm x}}^{(t)} \,\right]\,,
\quad  
X_{i,l}^{(t)}  \in  \langle 0, L \rangle\,. 
\label{equat:spat}
\end{equation}
The state of agent is characterized 
by the abstract vector of strategy
\begin{equation} 
{\bf S}_i^{(t)} = 
\left[\, S^{(t)}_{i,1}, 
S^{(t)}_{i,2}\,,\ldots S^{(t)}_{i,d_{\rm s}}\,\right]\,,
\qquad S_{i,k}^{(t)} \in \langle 0,1 
\rangle\,. 
\label{equat:strategy_vector}
\end{equation}
considered with normalization  
\begin{equation}
\|\,  {\bf S}_i^{(t)}\, 
\|  \equiv  \sqrt{\sum_{k=1}^{d_{\rm s}} 
\left(\, S^{(t)}_{i,k}\,  \right)^2}   
= 1 \,.
\label{equat:normalizacia}
\end{equation}
Here ${\bf S}^{(t)}_i$ defines the relative 
importance of the particular strategies, but the amplitude 
of their pursuing is determined by the mass 
(see in subsec.~\ref{subs:growth}).
The strategic vector is defined as an abstract 
information carrier which determines the strength 
of inter-agent interaction. In what follows, 
we use $d_{\rm x}=2$ and $d_{\rm s}=10$. 

\subsection{Growth mass rules} \label{subs:growth}

In analogy with logistic growth, the mass of each agent 
is considered to evolve according to discrete dynamics 
as it follows
\begin{equation}
M_i^{(t+1)} =  \alpha M_i^{(t)} -  \beta  \Omega_i^{(t)}\,.
\label{equat:rast}
\end{equation}
Here $\alpha$ is the constant growth rate parameter 
and $\beta$ the feedback parameter.  Generally, the second
term $(-\beta \Omega_i^{(t)})$ describes
the effect of pairwise competition. 
In further, we consider the regime with $\alpha>1$. 
In that case a discrete  model incorporates the growth 
property. 

The competitive term in Eq.(\ref{equat:rast}) is based on the 
overlap   
\begin{equation}
\Omega_i^{(t)}  
\equiv  \Omega({\bf X}_i^{(t)},{\bf S}_i^{(t)} )
= \sum_{i\neq j}^N \, J^{(t)}_{i,j} 
\sum_{k=1}^{d_{\rm s}}\, 
S_{i,k}^{(t)} S_{j,k}^{(t)}
\label{equat:omega}
\end{equation}
weighted by the pair-wise real-space distance matrix  
\begin{equation}
J_{i,j}^{(t)} 
=J\frac{M_i^{(t)}\, M_j^{(t)}    
}{
\left( \, \|\, {\bf X}_i^{(t)} -  {\bf X}_j^{(t)}\,\|^2
+\epsilon^2\,\right)^{\gamma/2}}\,,
\label{equat:jij}
\end{equation} 
which takes into account actual positions of agents. 
The tensorial structure $M_i^{(t)}\, M_j^{(t)}$     
has been chosen in analogy to scalar 
form of decay outlined by Eq.(\ref{equat:decay}).  
An important feature of agents is their ability to perform 
transformation of scalar nourishment $\alpha M_i$ 
into "vector mass" $M_i {\bf S}_i$ diversified 
by the components of ${\bf S}_i$.
In this context it has to be reminded that growing mass 
increases the impact of the respective strategy 
on the dynamics of its neighbors. Quite analogously 
as in the logistic function, we define the pairwise 
interaction to be proportional to the product 
of virtual masses and interaction 
parameter $J$. 
At the large inter-agent 
distances interaction turns 
to the asymptotics $\propto 
M_i^{(t)} M_j^{(t)} 
\| {\bf X}_i^{(t)} - {\bf X}_j^{(t)}\|^{-\gamma}$.
For the small distances, the parameter $\epsilon$ 
is introduced to prevent from the proximity 
effects.  The chosen form of the 
inter-agent pair interactions 
is of the short-range type 
($\gamma$, 
see in Tab.~\ref{tabl:values}), 
spherically symmetric and purely 
repulsive.

\subsection{Optimization as a individual coevolutionary mechanism in dynamic
landscape}\label{subs:local_optim}

The optimization starts by the analysis of external
stimuli represented by actual $\Omega_i$. In general, 
individual optimization is the coevolutionary 
mechanism which serves to adapt to local competitive 
dynamical environment~\cite{Wil01,Chan06} 
understood as formed by the surrounding agents. 

In our model, the optimization of $\Omega_i^{(t)} \equiv  
\Omega\left( {\bf X}_i^{(t)}, {\bf S}_i^{(t)}\right)$  
is considered to be the cause of the motion in the coordinate 
and strategic spaces. Here we use the {\em hill climbing} 
optimization technique which is the standard 
component of agent-based modeling~\cite{Chan06}. 
The expected consequence of the optimization is weakening  
of the competition pressure on $i$-th agent.  
According to Eq.(\ref{equat:rast}) 
smaller $\Omega_i$ leads to slower loss of the 
mass in onward iterations. 
The optimization of the selected agent 
is applied with probability $P_{\rm opt}$.  
Technically, in the simulation 
process the optimization 
is accepted if $P_{\rm opt}$  becomes larger 
than random number drawn uniformly from $(0,1)$.  
When the step is accepted, the agent has to 
decide among two alternatives: positional or 
strategy optimization. The individual optimization 
process which claims to find better position   
\begin{equation}
{\bf X}_i^{(t+1)} 
=  \hat{H} \left({\bf X}^{(t)}_i, N_{\rm H}, \delta_{\rm x}^{(t)} 
\right)
\label{equat:xH_vseobecne}
\end{equation}
is accepted with probability $P_{\rm ps}$. Here $\hat{H}$ 
denotes the hill climbing operator characterized by 
$N_{\rm H}$ variable displacements 
$\delta_{\rm x}^{(t)}$.  The iterations 
of strategic optimization 
formally written as  
\begin{equation}
{\bf S}_i^{(t+1)}
 =  \hat{H} 
\left(\, 
{\bf S}^{(t)}_i,  N_{\rm H}, 
\delta_{\rm s}^{(t)} \right) 
\end{equation}
are accepted with probability $1 - P_{\rm ps}$. 

At first, Let as focus on the spatial 
optimization in more details.
Formally, $\hat{H}$ applied to the search 
for optimal coordinates ${\bf X}_i$ has been 
decomposed into particular tasks 
solved by the sub-operators 
\begin{eqnarray} 
{\bf X}_{i,(n)}^{(t)} = 
\hat{H}_{\rm sub} \left( {\bf X}_{i,{(n-1)}}^{(t)}, \delta_{\rm x}^{(t)} \right)\,,
\qquad n=1, 2,  3,  \ldots,  N_{\rm H}\,, 
\label{equat:nH234}
\end{eqnarray} 
where ${\bf X}_{i,(0)}^{(t)} = {\bf X}_{i}^{(t)}$ 
represents initial 
coordinates for the optimization process. 
For each of $N_{\rm H}$ steps 
we perform the calculation of corresponding $\Omega_i$ 
for actual trial coordinates. 
It consists of calculation of 
$J_{i,j}$ matrix and overlap 
by means of Eq.(\ref{equat:omega}) and Eq.(\ref{equat:jij}). 
For $k$th spatial coordinate, and $n$th 
application of 
Eq.(\ref{equat:nH234}) we suppose the trial move  
\begin{equation}
X^{(t),{\rm trial}}_{i,k,(n)} = X^{(t)}_{i,k,(n-1)} 
+  \delta _{\rm x}^{(t)} \left( 2 \xi_{i,k,(n-1)}^{(t)} -1 \, \right)\,,
\label{equat:trial}
\end{equation} 
where $\xi^{(t)}_{i,k,(n-1)}$ is a random number drawn from a 
uniform 
distributions $(0,1)$.  The agent-based system 
is not conservative,  but its square boundaries 
are impenetrable by mass. The agents must 
restrict their moves within boundaries, 
and thus path corrections are needed 
for $X^{(t),{\rm trial}}_{i,k,(n)}$. 

The optimization uses two alternative displacements~\cite{Cap08}:
$\delta_{\rm x}^{(t)}  \in  \{ \delta_{{\rm x}_{1}},  \delta_{{\rm x}_{2}} 
\},$  $  \mbox{where}\, $   $ \delta_{\rm x_{1}}>  \delta_{\rm x_{2}}$.
The larger step $\delta_{{\rm x}_1}$  is drawn with probability 
$P_{\rm big}$.  The dichotomy of steps has intuitive 
reasons supported by the preliminary simulations. 
The step size $\delta_{{\rm x}_1}$ can be efficient 
to find location in the remote areas,  
whereas the displacement $\delta_{\rm x_{2}}$ 
helps to refine the spatial positions. 

The constant (spatially uniform) $\alpha$ in model does 
not guarantee automatically uniform access 
to external resources as the free boundary 
conditions are imposed. The character of interactions combined
with given conditions brings permanent heterogeneity in the access 
to external sources.  The nonequivalent  
persistent nonuniformities at corners and edges 
can be in particular overcome by considering 
sufficiently large systems 
and short-range interactions. 
Formally, the optimization consists of the sequence of particular 
decisions given by 
\begin{eqnarray}
\mbox{\large$\bf if$} 
&&  
\,\, \Omega\left({\bf X}_{i,(n)}^{(t),{\rm trial}},{\bf S}_i^{(t)}\right) 
\leq  
\Omega\left({\bf X}_{i,(n-1)}^{(t)},{\bf S}_i^{(t)}\right)
\\
& &\, \, \, \, \, \,\,  
\mbox{$\bf then$}
\,\,\,\,\,   
{\bf X}_{i,(n)}^{(t)} =  
{\bf X}_{i,(n)}^{(t),{\rm trial}}  
\,\,\,\,\,\,  
\mbox{$\bf otherwise $} 
\,\,\,\,
{\bf X}_{i,(n)}^{(t)}   =  
{{\bf X}_{i,(n-1)}^{(t)}}\, .
\nonumber
\end{eqnarray}
When $n = N_{\rm H}$ the output of individual optimization
${\bf X}^{(t+1)}_i= {\bf X}_{i,(N_{\rm H})}^{(t)}$  
is achieved [see Eq. (\ref{equat:xH_vseobecne})]. 

The analogous procedure which uses two different 
types of steps $\delta_{{\rm s}_1}$, 
$\delta_{{\rm s}_2}$ is assumed for 
the optimization of ${\bf S}_i$.  The only exception 
is the normalization of the strategy vector 
[see Eq.(\ref{equat:normalizacia})] which has to be applied 
after each optimization move.

The above presented agent-based model shares many
features with bacterial species optimizing 
their access to nutrients in different conditions~\cite{Pai03}. 
In this context the identification of $\Omega_i$ is analogous 
to output of bacterial chemosensory system and the decision from the optimization 
can be the switching between staying in the same or moving to better place.

\subsection {Mass constraints, birth/death processes} \label{subs:extrem} 

The iterative scheme Eq.(\ref{equat:rast}) fails to describe mass 
dynamics when passing to extremal values of masses.  
This deficiency of the model must be eliminated by additional 
constraints and limitations.  We have considered the mass 
bounded by the lower cutoff, $M_{\rm d}$. If, due to competition,  
the mass has decreased bellow $M_{\rm d}$, 
the agent is replaced by a new one in random initial state. 
The situation may be interpreted as death or crash. 
As the number of agents is conserved, 
the death of one agent opens the playground for immediate 
birth of his descendant with parameters drawn 
in a random way, analogously as in the stage of initialization
(i.e. with initial mass equal to $2 \,M_{\rm d})$.
The operating near threshold $M_{\rm d}$ 
yields ergodicity gain in a way of 
extremal dynamics~\cite{Bak88}, which avoids 
from getting stuck.

Preliminary numerical simulations uncovered 
that there remains specific issue unresolved
related to nonstationarity of mass distributions ~\cite{Wit05}
caused by the growth very weighty agents.  
More profound analysis showed that quadratic form $M_i M_j$ involved in Eq.(\ref{equat:jij}) 
might not stop the growth if the masses of competing 
neighbours are not sufficiently high. 
Stationarity may be reached by introducing upper mass cutoff 
$M_{\rm up}$. By the update $M_i^{(t)}  \leftarrow  M_{\rm up}$ 
the agent suddenly reacts 
to the situation $M_i^{(t)} > M_{\rm up}$.  
This limitation can be understood as an extra 
constraint which guarantees renewability of sources. 
In the economic context the 
introducing of $M_{\rm up}$ 
may roughly represent very restrictive 
taxation system.

\begin{table} 
\centering
\begin{tabular}{|c|c|c|c|} 
\hline
Symbol & Meaning &  Value & Introduced \\
\hline
$\alpha$ & {\scriptsize constant growth parameter} & $1.2$ &  Eq.\ref{equat:rast}\\
$\beta$ & {\scriptsize feedback parameter} & varying &  Eq.\ref{equat:rast} \\
$\gamma$ & {\scriptsize exponent of interaction} & $4$ &  Eq.\ref{equat:jij}\\
$\epsilon$ & {\scriptsize parameter of interaction} & $0.0005$ &
Eq.\ref{equat:jij}\\
$N$ & {\scriptsize number of agents} & $400$ &  subsec.\ref{subs:stateof}\\
$d_{\rm x}$ & {\scriptsize dimension of spatial coordinate} & $2$ & 
Eq.\ref{equat:spat}\\
$d_{\rm s}$ & {\scriptsize dimension of strategic variable} & $10$ &
Eq.\ref{equat:strategy_vector}\\
$\delta_{\rm x_{1}}$ & {\scriptsize small step of spatial optimization} & $0.01$ &  
subsec.
\ref{subs:local_optim}\\
$\delta_{\rm x_{2}}$ & {\scriptsize long step of spatial optimization} & $0.2$ &  
subsec.\ref{subs:local_optim}\\
$\delta_{{\rm s}_1}$ & {\scriptsize small step of strategy optimization} & $0.003$ &  
subsec.\ref{subs:local_optim}\\
$\delta_{{\rm s}_2}$ & {\scriptsize long step of strategy optimization} & $0.03$ &  
subsec.\ref{subs:local_optim}\\
$J$ & {\scriptsize interaction strength controller} & $1$ &  
Eq.\ref{equat:jij}\\
$L$ & {\scriptsize square segment length} & $4$ &  
Eq.\ref{equat:spat}\\
$M_{\rm d}$ & {\scriptsize lower threshold for agent mass} & $0.02$ &  
subsec.\ref{subs:extrem}\\
$M_{\rm up}$ & {\scriptsize upper threshold for agent mass} & $100$ &  
subsec.\ref{subs:extrem}\\
$P_{\rm big}$ & {\scriptsize selection probability of bigger steps} & $0.2$ &  
subsec.\ref{subs:local_optim}\\
$P_{\rm ps}$ & {\scriptsize probability of decision to optimize} & $0.3$ &  
subsec.\ref{subs:local_optim}\\
\hline
\end{tabular}
\caption{Numerical values of model parameters 
used in the simulations supplemented 
by the short explanation and link 
to the main text where the issue is 
introduced.}
\label{tabl:values}
\end{table}

\section{Results}\label{sec:secnum1}

During the assembly preparation, the positions ${\bf X}_i$  
and as well as vectors of strategies ${\bf S}_i$ are initialized by 
random values and initial values of masses is taken to be 
$2 M_{\rm d}$. Simulations have been carried out 
for the values of parameters listed 
in Table~\ref{tabl:values}. 

\begin{figure}[!ht]
\centering
\includegraphics[width=13cm]{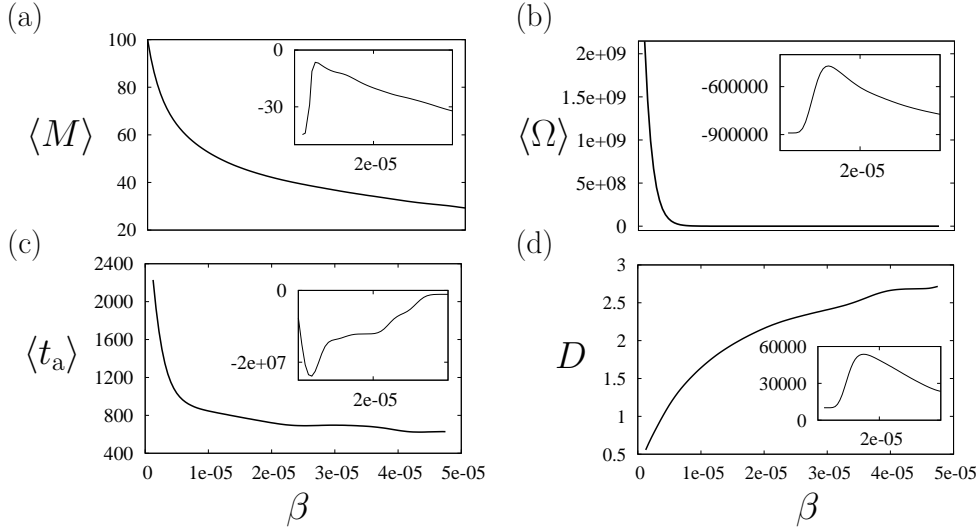}
\caption{
The mean equilibrium characteristics in LCP and HCP 
phases plotted as a functions of the parameter $\beta$. 
The partial figures depict information 
about the mean:  (a)~{\em agent's mass} $\langle M \rangle$; 
(b)~{\em mutual overlap} $\langle \Omega \rangle$ 
(match of strategies weighted by distance of agents);
(c)~{\em life expectancy} $\langle t_a \rangle$ 
reduced by competition at 
high $\beta$;  (d)~{\em diffusion coefficient} 
$D$ showing enhanced migration effect when 
the competition gets more intense. 
Each part of figure is supplemented 
by the inset showing how the first derivative 
of the corresponding quantity changes with respect to $\beta$.  
All the anomalies are localized near the expected 
critical point. The main drawbacks here are the differences 
caused by the finite size 
effects. The dependences have been obtained by averaging 
of 13 independent runs going from the 
low to high $\beta$.  For each run and 
each fixed $\beta$ we treated record of data corresponding to 
40 000 random visits per agent.}
\label{fig:obrc}
\end{figure}

One of the main purposes of the paper is to understand 
the impact of competition to the system dynamics.
We decided to construct $\beta$-dependences of the mean 
statistical values. We start by rather artificial regime $\beta=0$.
In this trivially noncompetitive situation all masses attain 
$M_{\rm up}$ and optimization is equivalent to random walk. 
In Fig.(\ref{fig:obrc})(a) we depict $\beta$-dependent effect  
of the competitive reduction of the mean mass 
$\langle M \rangle$, where $\langle \dots \rangle$ stands for 
the numerical averaging over the time and 
assembly of the agents.

\begin{figure}[!ht]
\centering
\includegraphics[width=13cm]{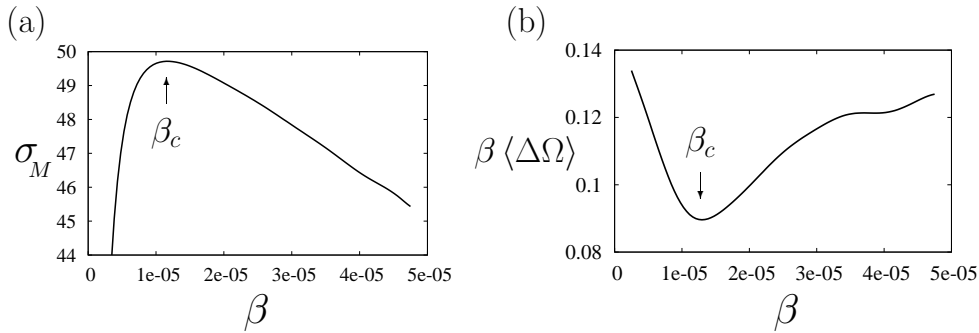}
\caption{
The study of the fluctuations and their anomalies. 
The parts of figure include: 
(a)~{\em mass dispersion} $\sigma_{\rm M}$;  
(b)~the measure of benefits $\langle \Delta \Omega \rangle$
that are gained by the short-time optimization.
The extremes serve to identify 
critical point $\beta_{\rm c}$.}
\label{fig:obrsig}
\end{figure}

More profound picture of the system behavior is achieved 
by analyzing the statistics of mass fluctuations.  The mass 
fluctuations around the mean $\langle M \rangle $ 
are characterized by the mass dispersion 
$\sigma^2_{\rm M} = \langle M^{2} \rangle  - 
\langle M \rangle^{2}$ [see Fig.\ref{fig:obrsig}(a)].
The dependence of 
$\sigma^{2}_{\rm M}$ 
exhibits extreme at $\beta = \beta_{\rm c} = 1.2 \times 10^{-5}$.
The value may be interpreted as the {\em critical point} 
of the phase transition driven by $\beta$.  
This scenario is consistent with the behaviour 
of the corresponding derivatives shown in insets of 
Fig.(\ref{fig:obrc}). The critical point separates 
phases of different competition (HCP and LCP, respectively). 
The phases were named according behavior of 
$\langle \Omega \rangle$ [Fig.\ref{fig:obrc}(b)]. 
For higher $\beta$ we observed better
(better in a sense of smaller $\langle \Omega\rangle$) 
way to minimize $\Omega_i$ that agents of HCP become 
self-improved because of more carefully 
chosen strategies. 

\begin{figure}[!ht]
\centering
\includegraphics[width=13cm]{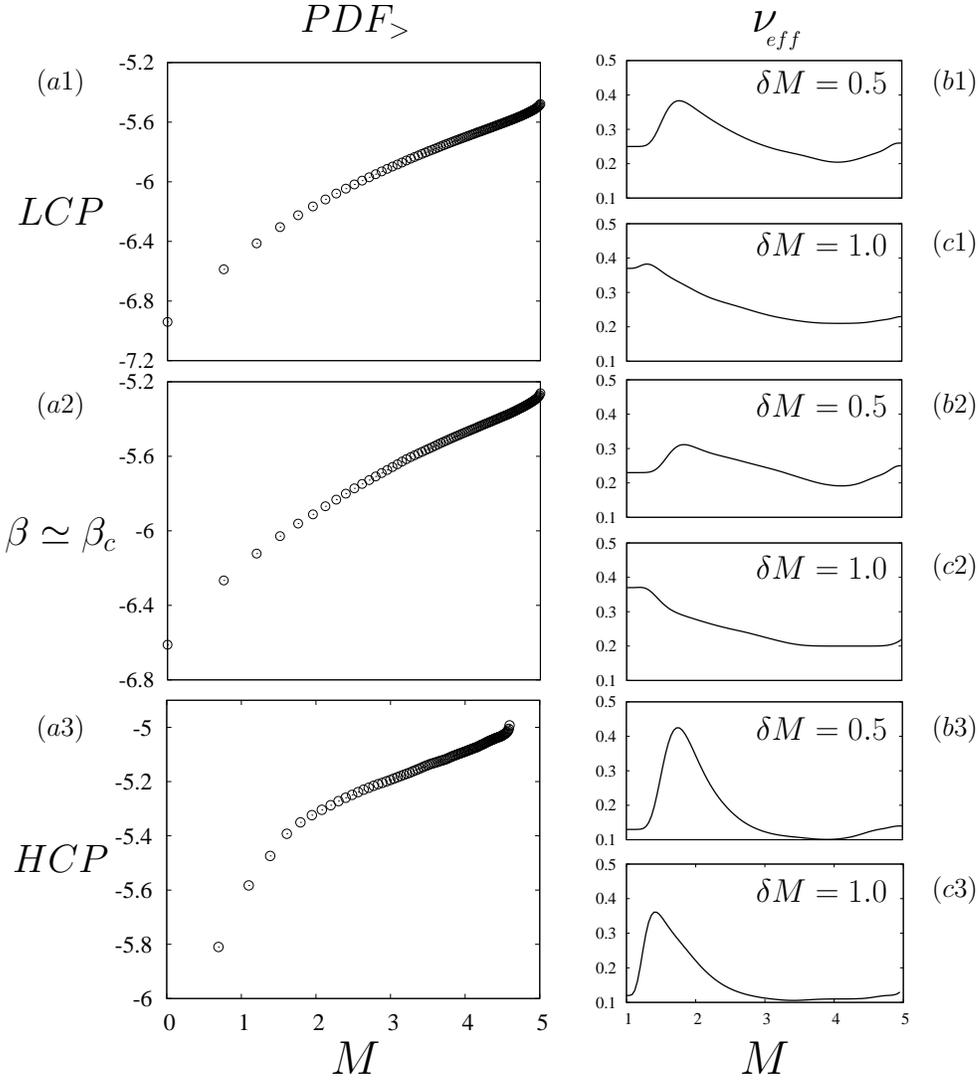}
\caption{  
The cumulative probability 
distributions and corresponding 
effective exponents (right hand side plots). 
Calculated for simulation data 
obtained for three representative 
values of $\beta$:
(a1)~$\beta = \beta_{\rm c} / 10$, 
(a2)~$\beta = \beta_{\rm c}$ and 
(a3)~$\beta =10 \beta_{\rm c}$. 
The local properties of distributions characterized 
by the effective exponent $\nu_{\rm eff}$ which is a 
function of $M$ 
($M$ is always middle point of the local fit) for 
two different resolutions $\delta M$. For 
$\delta M=0.5$ and corresponding 
$\beta$ we found that:  
(b1) $\nu_{\rm eff}\in (0.2,0.38)$; 
(b2) $\nu_{\rm eff}\in (0.2,0.31)$; 
(b3) $\nu_{\rm eff}\in (0.1,0.42)$. 
It means that the lowest spread of  local effective 
exponent (b2) 
corresponds to the critical regime.}
\label{fig:cdf}
\end{figure}

The power-law distributions are assumed to be the 
hallmarks of criticality~\cite{New05}.  
It is therefore instructive to identify the specific features of distributions 
regarding position of critical point. 
The mass distributions have been studied 
for three representative $\beta$ values:
$\beta = 0.1  \beta_{\rm c}$,  $\beta_{\rm c}$, 
$10 \beta_{\rm c}$ [see Fig. (\ref{fig:cdf})]. 
Despite significant distance from the critical 
regime we see that distributions may be roughly 
characterized by the power-law distributions with 
anomalies concentrated around the tails of high 
and low mass regions.  The parametric robustness 
of the power-law like distributions may be 
considered as realistic feature of the model.  
It is reasonable to suppose that anomalies 
originate prevailingly from the mass constraints 
(see subsec.~\ref{subs:extrem}).  
In our relatively small-size system the transition 
is not sharp but broad in $\beta$. Also performed detailed 
analysis of exponents ensures
parametric robustness. To characterize the distributions 
more quantitatively, the respective $M$ dependence 
of effective local index  
$\nu_{\rm eff}$$(M, \delta M)$ has been calculated for each of them.
These dependences have been 
constructed by fitting of power-law functions
of some exponent and amplitude on $(- \delta M + M ,  M +  \delta M)$ for 
varying central $M$. Any plateau of $\nu_{\rm eff}(M,\delta M)$  
indicates the existence of particular 
power-law behavior 
on the domains larger than $\delta M$.
It should be noted that our idea of usage of local 
fits has been motivated by the phenomenology~\cite{Coe08,Sca07}.
From the results obtained for different $\beta$ 
we see that formation of power-law ditribution 
exhibits strong robustness with respect 
to $\beta_{\rm c}$ in the sense that particular 
local intervals of the power-law behavior stay 
sufficiently far from $\beta_{\rm c}$.  On the other hand, 
the classical concept of the 
phase transition expresses 
itself through minimum spread 
of $\nu_{\rm eff}\simeq  0.19 - 0.31$ 
at the critical point. 

In the following we supplement our optimization 
results with the information about transport properties. 
The diffusion coefficient is an indirect characteristics 
which is related to the spatial individual 
optimization. 
Its definition  
\begin{eqnarray}
D =  
 \frac{1}{N}
  \sum_{i=1}^N
\langle \, 
  \parallel  {\bf X}_{i}^{ (t+\tau)} 
-   {\bf X}_i^{(t)} 
 \parallel^2\, 
 \rangle_{(\tau)}
 \end{eqnarray}
captures space and time averaging of pathways. 
The index $(\tau)$ used to highlight
data has been averaged over the 
time scale on the sub-assembly of the agents 
which live longer than $\tau$.  
The above mentioned averages are depicted 
in Fig.(\ref{fig:obrc})(d). 
They indicate that HCP agents spread 
faster owing to reduced interactions, 
caused by their reduced mass, 
abating blocking from the rivals. 

The short-time efficiency of the optimization processes
is analyzed using the fitness~\cite{Beveridge} 
function defined by the difference
\begin{equation}
\Delta \Omega^{(t)}_i =
\Omega({\bf X}_{i,(N_{\rm H})}^{(t)} , {\bf S}_i^{(t)}) -
\Omega({{\bf X}_{i,(0)}^{(t)}} , {\bf S}_i^{(t)})\,.
\end{equation}
(The same differences have been calculated and averaged when 
the optimization of strategy vectors is considered.) According to the 
above measure, the optimization of the position or strategy 
is more efficient if the difference becomes higher. 
But the difference measure tells little about the 
long-time perspectives of the agents. 
Namely at HCP the landscape varies very rapidly 
and  optimization yields only the short-time 
benefits from the movement. In the conditions of 
critical regime [see Fig.(\ref{fig:obrsig}) (b)] 
the mean $\langle \Delta \Omega \rangle$ 
shows minimum. This confirms known fact that 
critical landscapes are landscapes 
of the extremal 
complexity~\cite{Lan90,Chan06}.  

The life expectancy characteristics depicted in 
Fig.(\ref{fig:obrc})(c) indicates that the mean 
lifespan of HCP agents is considerably shorter 
in comparison with LCP agents. The combination of the 
above facts clearly indicates how inefficient the short-time 
optimization without prediction could be, namely in 
HCP constrictive conditions. The effect of enhanced  
diffusion as well as the effect of shortened life 
expectancy of agents could be considered as a typical 
examples of emergence, which has not been 
a priory integrated into 
the architecture of agents.

\begin{figure}[!ht]
\centering
\includegraphics[width=13cm]{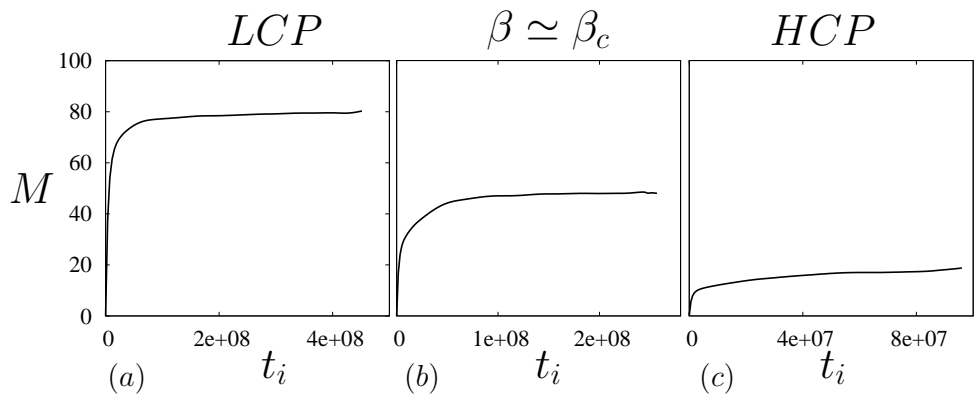}
\caption{
The return to primary 
motivation for the model construction 
[see Eq.(\protect\ref{equat:logis1})].  
The parts showing the individual mean mass growth as a 
function of the lifespan of agents.  
Each age group is averaged separately. 
Growth dependence is calculated for 
three representative values 
of $\beta$ corresponding to: 
(a)~LCP phase, ($\beta=\beta_{\rm c}/10$); 
(b)~critical regime; 
(c)~HCP phase, ($\beta=10 \beta_{\rm c}$).}
\label{fig:logistic}
\end{figure}

Finally our study has been focused on the mean individual growth 
of the agent as a function of his lifespan. The dependences
depicted in Fig.(\ref{fig:logistic}) demonstrate 
saturation in qualitative agreement with the previously 
mentioned logistic forms. The analysis shows that despite 
the typically nonequilibrium nature of growth models, 
due to the projection of the mass into
individual lifespan coordinate (and subsequent averaging
over the assembly of masses belonging to the respective age), 
the mean growth dependences may be identified 
in the steady-state.   For such numerical output, 
the carrying capacity can be identified a posteriori. 
Since agents often imitate mutually destructive and self-destructive 
actions, it seems rather surprising that saturation 
without recession appears at large 
lifespans.  The qualitative microexplanation 
is that the drop of the mass of some agents 
is often accompanied by empowered growth of their 
respective neighbors and thus finally 
low-rate growth stem from the most of 
the rivalry crowds. 



\clearpage

\section{Conclusion}

In the paper the agent-based model of competitive behavior
with implemented procedure of individual optimization was
investigated. Our study focused on the equilibrium 
statistics and efficiency of the individual optimization.  
The principal finding is that there exists 
critical regime which indicates transition 
from the LCP to HCP phase and that there are significant
differences in the efficiency of optimization in the
respective phases. In HCP the low order resources 
are turned to the strategically well organized matter. 
The anomaly in the  efficiency belongs to the complex 
barriers of $\Omega_i$ corresponding to critical point. 
We observed nearly power-law behavior of the mass distributions
robust with respect to parametric choices.   
The focus on the 
tails of mass distributions suggests nonequivalence 
of central and close to boundary space positions, 
which cause nonuniform access to the 
external sources.  

Among the emergent phenomena, 
which typically accompany the agent-based 
simulations, we could mention higher mobility 
of lighter agents and lifespan reduced 
by their motion close to lower mass region.  
In the future studies we plan to investigate 
the impact of an extra payoffs 
for the optimization and mobility 
which may strengthen competetivness.  

In the paper we present results of the equilibrium simulations 
of the growth. As the equilibrium conditions are not always
suitable for the growth problems, further perspectives of given model 
can be seen in nonequilibrium applications 
(e.g. in the models of metastatic growth with dissemination of malignant
cells). It would be also interesting 
to combine individualized distributed parameters, 
e.g. those for decision to optimize (agent-dependent, 
distributed $P_{{\rm ps},i}$
instead of uniform $P_{\rm ps}$), and analyze their impact 
on the mass statistics.  Further perspectives can  be seen 
in the application of the realistic geografic 
boundary conditions, space-distributed sources 
$\alpha( {\bf x} )$ and assortment related to the 
specific sources. 

\noindent{\bf Acknowledgements}

The authors 
acknowledge financial support 
from VEGA, Slovak Republic 
(Grant No. 1/4021/07). One of the authors,
D.H. acknowledges ﬁnancial support by a postdoc-                             ´
toral fellowship {\sc LeStudium} of the Region Centre and the
Centre National de la Recherche Scientiﬁque
(during his stay at the CBM and MAPMO CNRS institutes).

\newpage

\bibliographystyle{elsart-num}
\bibliography{bibi2}

\end{document}